\documentclass[runningheads]{llncs}
\usepackage{amssymb}
\usepackage{amsmath}
\usepackage{graphicx}
\usepackage{acronym}
\usepackage{theorem}
\usepackage{multirow}
\usepackage{verbatim}
\usepackage{booktabs}
\usepackage{color}
\usepackage{algorithm}
\usepackage{algpseudocode}
\usepackage{pifont}
\newcommand{\xmark}{\ding{53}}%

\algdef{SE}[DOWHILE]{Do}{doWhile}{\algorithmicdo}[1]{\algorithmicwhile\ #1}%
\usepackage{enumitem}
\usepackage{url}
\urldef{\mailsa}\path|p.santini@pm.univpm.it,m.baldi@univpm.it, f.chiaraluce@univpm.it|
\newcommand{\keywords}[1]{\par\addvspace\baselineskip
\noindent\keywordname\enspace\ignorespaces#1}

{\theorembodyfont{\rmfamily}
   \newtheorem{The}{{\textbf Theorem}}[section]}
{\theorembodyfont{\rmfamily}
   }
{\theorembodyfont{\rmfamily}
   }
{\theorembodyfont{\rmfamily}
   }
{\theorembodyfont{\rmfamily}
   }

\acrodef{LDPC}{low-density parity-check}
\acrodef{SDP}{syndrome decoding problem}
\acrodef{LDGM}{low-density generator matrix}
\acrodef{MDPC}{moderate-density parity-check}
\acrodef{QC}{quasi-cyclic}
\acrodef{QC-LDPC}{quasi-cyclic low-density parity-check}
\acrodef{QC-MDPC}{quasi-cyclic moderate-density parity-check}
\acrodef{RSA}{Rivest-Shamir-Adleman}
\acrodef{BF}{bit flipping}
\acrodef{SPA}{sum product algorithm}
\acrodef{RDF}{random difference families}
\acrodef{ISD}{information set decoding}
\acrodef{KRA}{key recovery attack}
\acrodefplural{KRA}[KRAs]{key recovery attacks}
\acrodef{DA}{decoding attack}
\acrodefplural{DA}[DAs]{decoding attacks}
\acrodef{WF}{work factor}
\acrodef{BER}{bit error rate}
\acrodef{CER}{codeword error rate}
\acrodef{BSC}{binary symmetric channel}
\acrodef{BPSK}{binary phase shift keying}
\acrodef{$2$-PAM}{binary pulse amplitude modulation}
\acrodef{AWGN}{additive white Gaussian noise}
\acrodef{LLR}{log likelihood ratio}
\acrodefplural{LLR}[LLRs]{log likelihood ratios}
\acrodef{SPA}{sum-product algorithm}
\acrodef{DFR}{decryption failure rate}
\acrodef{SL}{security level}
\acrodef{ECC}{elliptic curve cryptography}
\acrodef{QD}{quasi-dyadic}
\acrodef{GRS}{generalized Reed-Solomon}
\acrodef{DSA}{Digital Signature Algorithm}
\acrodef{ECDSA}{Elliptic Curve Digital Signature Algorithm}
\acrodef{KEM}{key encapsulation mechanism}
\acrodefplural{KEM}[KEMs]{key encapsulation mechanisms}
\acrodef{PKC}{public-key cryptosystem}
\acrodef{SK}{secret key}
\acrodef{PK}{public key}
\acrodef{CCA}{chosen ciphertext attack}
\acrodef{CCA2}{adaptive chosen ciphertext attack}
\acrodef{CPA}{chosen plaintext attack}
\acrodef{KI}{Kobara-Imai}
\acrodef{PFS}{Perfect Forward Secrecy}
\acrodef{NP}{nondeterministic-polynomial}
\acrodef{DRBG}{deterministic random bit generator}
\acrodef{TRNG}{true random number generator}
\acrodef{KDF}{key derivation function}
\acrodef{AKE}{authenticated key exchange}
\acrodef{KAT}{Known Answer Tests}

\begin{document}

\mainmatter  
\title{Assessing and countering reaction attacks against post-quantum public-key cryptosystems based on QC-LDPC codes}

\titlerunning{Analysis of reaction attacks against QC-LDPC code-based cryptosystems}

\author{Paolo Santini$^*$ \and Marco Baldi \and Franco Chiaraluce}
\authorrunning{P. Santini \and M. Baldi \and F. Chiaraluce}

\institute{Universit\`a Politecnica delle Marche, Ancona, Italy\\\mailsa}

\maketitle

\begin{abstract}
Code-based public-key cryptosystems based on QC-LDPC and QC-MDPC codes are promising post-quantum candidates to replace quantum vulnerable classical alternatives.
However, a new type of attacks based on Bob's reactions have recently been introduced and appear to significantly reduce the length of the life of any keypair used in these systems.
In this paper we estimate the complexity of all known reaction attacks against QC-LDPC and QC-MDPC code-based variants of the McEliece cryptosystem.
We also show how the structure of the secret key and, in particular, the secret code rate affect the complexity of these attacks.
It follows from our results that QC-LDPC code-based systems can indeed withstand reaction attacks, on condition that some specific decoding algorithms are used and the secret code has a sufficiently high rate.
\end{abstract}

\keywords{Code-based cryptography, McEliece cryptosystem, Niederreiter 
cryptosystem, post-quantum cryptography, quasi-cyclic low-density parity-check codes.}

\section{Introduction}
\let\thefootnote\relax\footnotetext{$^*$The work of Paolo Santini was partially supported by Namirial S.p.A.}

Research in the area of post-quantum cryptography, that is, the design of cryptographic primitives able to withstand attacks based on quantum computers has known a dramatic acceleration in recent years, also due to the ongoing NIST standardization initiative of post-quantum cryptosystems \cite{NISTcall2016}.
In this scenario, one of the most promising candidates is represented by code-based cryptosystems, that were initiated by McEliece in 1978 \cite{McEliece1978}. 
Security of the McEliece cryptosystem relies on the hardness of decoding a random linear code: a common instance of this problem is known as \ac{SDP} and no polynomial-time algorithm exists for its solution \cite{Berlekamp1978,May2011}.
In particular, the best \ac{SDP} solvers are known as \ac{ISD} algorithms \cite{Prange1962,Stern1989,Becker2012}, and are characterized by an exponential complexity, even considering attackers provided with quantum computers \cite{Bernstein2010}.

Despite these security properties, a large-scale adoption of the McEliece cryptosystem has not occurred in the past, mostly due to the large size of its public keys: 
in the original proposal, the public key is the generator matrix of a Goppa code with length $1024$ and dimension $524$, requiring more than $67$ kB of memory for being stored.
Replacing Goppa codes with other families of more structured codes may lead to a reduction in the public key size, but at the same time might endanger the system security because of such an additional structure. An overview of these variants can be found in \cite{Baldi2017}.

Among these variants, a prominent role is played by those exploiting \ac{QC-LDPC} \cite{Baldi2008,Baldi2012} and \ac{QC-MDPC} codes \cite{Misoczki2013} as private codes, because of their very compact public keys.
Some of these variants are also at the basis of post-quantum primitives that are currently under review for possible standardization by NIST \cite{BIKE,LEDAkem}. 
\ac{QC-LDPC} and \ac{QC-MDPC} codes are decoded through iterative algorithms that are characterized by a non-zero \ac{DFR}, differently from classical bounded distance decoders used for Goppa codes.
The values of \ac{DFR} achieved by these decoders are usually very small (in the order of $10^{-6}$ or less), but are bounded away from zero.

In the event of a decoding failure, Bob must acknowledge Alice in order to let her encrypt again the plaintext.
It has recently been shown that the occurrence of these events might be exploited by an opponent to recover information about the secret key \cite{Guo2016,Fabsic2017,Fabsic2018}.
Attacks of this type are known as \textit{reaction attacks}, and exploit the information leakage associated to the dependence of the \ac{DFR} on the error vector used during encryption and the structure of the private key.
These attacks have been shown to be successful against some cryptosystems based on \ac{QC-LDPC} and \ac{QC-MDPC} codes, but their complexity has not been assessed yet, to the best of our knowledge.

In this paper, we consider all known reaction attacks against \ac{QC-LDPC} and \ac{QC-MDPC} code-based systems, and provide closed form expressions for their complexity.
Based on this analysis, we devise some instances of \ac{QC-LDPC} code-based systems that are able to withstand these attacks.
The paper is organized as follows.
In Section \ref{sec:descryption} we give a description of the \ac{QC-LDPC} and \ac{QC-MDPC} code-based McEliece cryptosystems.
In Section \ref{sec:reaction} we describe known reaction attacks. 
In particular, we generalize existing procedures, applying them to codes having whichever parameters and take the code structure into account, with the aim to provide complexity estimations for the attacks.
In Section \ref{sec:further_cons} we make a comparison between all the analyzed attacks, and consider the impact of the decoder on the feasibility of some attacks.
We show that \ac{QC-LDPC} McEliece code-based cryptosystems have an intrinsic resistance to reaction attacks. 
This is due to the presence of a secret transformation matrix that implies Bob to decode an error pattern that is different from the one used during encryption.
When the system parameters are properly chosen, recovering the secret key can hence become computationally unfeasible for an opponent.

\section{System description\label{sec:descryption}}
Public-key cryptosystems and key encapsulation mechanisms based on \ac{QC-LDPC} codes \cite{Baldi2008,LEDAkem} are built upon a secret \ac{QC-LDPC} code with length $n=n_0 p$ and dimension $k = (n_0-1)p$, with $n_0$ being a small integer and $p$ being a prime. The latter choice is recommended to avoid reductions in the security level due to the applicability of folding attacks of the type in \cite{Shooshtari2016}.
The code is described through a parity-check matrix in the form:
\begin{equation}
\label{eq:privateH}
H=\left[ H_0 \left| H_1 \right| \cdots \left| H_{n_0-1} \right. \right],
\end{equation}
where each block $H_i$ is a $p\times p$ circulant matrix, with weight equal to $d_v$.

\subsection{Key generation\label{key:gen}}

The private key is formed by $H$ and by a transformation matrix $Q$, which is an $n \times n$ matrix in \ac{QC} form (i.e., it is formed by $n_0 \times n_0$ circulant blocks of size $p$). 
The row and column weights of $Q$ are constant and equal to $m \ll n$.
The matrix $Q$ is generated according to the following rules:
\begin{itemize}
\item the weights of the circulant blocks forming $Q$ can be written in an $n_0 \times n_0$ circulant matrix $w(Q)$ whose first row is $\bar{m}=\left[ m_0 , m_1 , \cdots , m_{n_0-1}\right]$, such that $\sum_{i=0}^{n_0-1}{m_i}=m$; the weight of the $(i,j)$-th block in $Q$ corresponds to the $(i,j)$-th element of $w(Q)$;
\item the permanent of $w(Q)$ must be odd for the non-singularity of $Q$; if it is also $<p$, then $Q$ is surely non-singular.
\end{itemize}
In order to obtain the public key from the private key, we first compute the matrix $\tilde{H}$ as:
\begin{equation}
\label{eq:Htilde}
\tilde{H}=HQ=\left[ \tilde{H}_0 \left| \tilde{H}_1 \right| \cdots \left| \tilde{H}_{n_0-1} \right. \right],
\end{equation}
from which the public key is obtained as:
\begin{equation}
\label{eq:public_key}
G'=\begin{bmatrix} I_{(n_0-1)p}  \left| \begin{matrix} \left(\tilde{H}^{-1}_{n_0-1} \tilde{H}_{0}\right)^T \\
\left(\tilde{H}^{-1}_{n_0-1} \tilde{H}_{1}\right)^T \\
\vdots \\
\left(\tilde{H}^{-1}_{n_0-1} \tilde{H}_{n_0-2}\right)^T \\ 
\end{matrix}\right.
\end{bmatrix},
\end{equation}
where $I_{(n_0-1)p}$ is the identity matrix of size $(n_0-1)p$.
The matrix $G'$ is the generator matrix of the public code and can be in systematic form since we suppose that a suitable conversion is adopted to achieve indistinguishability under \ac{CCA2} \cite{Kobara2001}.

\subsection{Encryption}

Let $u$ be a $k$-bit information message to be encrypted, and let $e$ be an $n$-bit intentional error vector with weight $t$. The ciphertext $x$ is then obtained as:
\begin{equation}
\label{eq:enc_x}
x=uG'+e.
\end{equation}
When a \ac{CCA2} conversion is used, the error vector is obtained as a deterministic transformation of a string resulting from certain public operations, including one-way functions (like hash functions), that involve the plaintext and some randomness generated during encryption.
Since the same relationships are used by Bob to check the integrity of the received message, in the case with \ac{CCA2} conversion performing an arbitrary modification of the error vector in \eqref{eq:enc_x} is not possible.
Analogously, choosing an error vector and computing a consistent plaintext is not possible, because it would require inverting a hash function. 
As we will see next, this affects reaction attacks, since it implies that the error vector cannot be freely chosen by an opponent.
Basically, this turns out into the following simple criterion:
in the case with \ac{CCA2} conversion, the error vector used for each encryption has to be considered as a randomly extracted $n$-tuple of weight $t$.

\subsection{Decryption}

Decryption starts with the computation of the syndrome as:
\begin{equation}
\label{eq:dec_syn}
s=xQ^TH^T=eQ^TH^T=e'H^T,
\end{equation}
which corresponds to the syndrome of an \textit{expanded error vector} $e'=eQ^T$, computed through $H^T$.
Then, a syndrome decoding algorithm is applied to $s$, in order to recover $e$. 
A common choice to decode $s$ is the \ac{BF} decoder, firstly introduced in \cite{Gallager1963}, or one of its variants.
In the special setting used in \ac{QC-LDPC} code-based systems, decoding can also be performed through a special algorithm named \textit{Q-decoder} \cite{LEDAkem}, which is a modified version of the classical \ac{BF} decoder and exploits the fact that $e'$ is obtained as the sum of rows from $Q^T$.  
The choice of the decoder might strongly influence the probability of success of reaction attacks, as it will be discussed afterwards.

\ac{QC-MDPC} code-based systems introduced in \cite{Misoczki2013} can be seen as a particular case of the \ac{QC-LDPC} code-based scheme, corresponding to $Q=I_{n_0p}$.
Encryption and decryption work in the same way, and syndrome decoding is performed through \ac{BF}. 
We point out that the classical \ac{BF} decoder can be considered as a particular case of the Q-decoder, corresponding to $Q=I_{n_0p}$.

\subsection{Q-decoder \label{sec:Qdecoder}}

The novelty of the Q-decoder, with respect to the classical \ac{BF} decoder, is in the fact that it exploits the knowledge of the matrix $Q$ to improve the decoding performance.
A detailed description of the Q-decoder can be found in \cite{LEDAkem}.
In the Q-decoder, decisions about error positions are taken on the basis of some \textit{correlation} values that are computed as:
\begin{equation}
R = s* H * Q = \Sigma * Q,
\end{equation}
where $*$ denotes the integer inner product and $\Sigma = s*H$.
In a classical BF decoder, the metric used for the reliability of the bits is only based on $\Sigma$, which is a vector collecting the number of unsatisfied parity-check equations per position.
In \ac{QC-LDPC} code-based systems, the syndrome $s$ corresponds to the syndrome of an \textit{expanded error vector} $e' = e Q^T$: this fact means that the error positions in $e'$ are not uniformly distributed, because they depend on $Q$.
The Q-decoder takes into account this fact through the integer multiplication by $Q$ \cite[section 2.3]{LEDAkem}, and the vector $R$ is used to estimate the error positions in $e$ (instead of $e'$).

In the case of \ac{QC-MDPC} codes, a classical \ac{BF} decoder is used, and it can be seen as a special instance of the Q-decoder, corresponding to $Q=I_{n_0p}$.
As explained in \cite[section 2.5]{LEDAkem}, from the performance standpoint the Q-decoder approximates a \ac{BF} decoder working on $\tilde{H} = HQ$. However, by exploiting $H$ and $Q$ separately, the Q-decoder achieves lower complexity than \ac{BF} decoding working on $\tilde{H}$.
The aforementioned performance equivalence is motivated by the following relation:
\begin{align}
\label{eq:correlation_Q}
R & = s * H * Q =\nonumber \\
& = eQ^TH^T*H*Q=\nonumber\\
& = e\tilde{H}^T * H *Q \approx\nonumber \\
& \approx e\tilde{H}^T * \tilde{H},
\end{align}
where the approximation $H Q \approx H * Q$ comes from the sparsity of both $H$ and $Q$.
Thus, equation \eqref{eq:correlation_Q} shows how the decision metric considered in the Q-decoder approximates that used in a \ac{BF} decoder working on $\tilde{H}$.

\section{Reaction attacks\label{sec:reaction}}
In order to describe recent reaction attacks proposed in \cite{Fabsic2017,Fabsic2018,Guo2016}, let us introduce the following notation.

Given two ones at positions $j_1$ and $j_2$ in the same row of a circulant block, the distance between them is defined as $\delta(j_1,j_2) = \min\left\{\pm(j_1-j_2) \mod{p} \right\}$.
Given a vector $v$, we define its distance spectrum $\Delta(v)$ as the set of all distances between any couple of ones.
The multiplicity $\mu^{(v)}_d$ of a distance $d$ is equal to the number of distinct couples of ones producing that distance; if a distance does not appear in the distance spectrum of $v$, we say that it has zero multiplicity (i.e., $\mu^{(v)}_d = 0$), with respect to that distance.
Since the distance spectrum is invariant to cyclic shifts, all the rows of a circulant matrix share the same distance spectrum; thus, we can define the distance spectrum of a circulant matrix as the distance spectrum of any of its rows (the first one for the sake of convenience).

The main intuition behind reaction attacks is the fact that the \ac{DFR} depends on the correlation between the distances in the error vector used during encryption and those in $H$ and $Q$.
In fact, common distances produce cancellations of ones in the syndrome, and this affects the decoding procedure \cite{Eaton2018}, by slightly reducing the \ac{DFR}.
In general terms, a reaction attack is based on the following stages:
\begin{enumerate}
\item The opponent sends $T$ queries to a decryption oracle. For the $i$-th query, the opponent records the error vector used for encryption ($e^{(i)}$) and the corresponding oracle's answer ($\Im^{(i)}$). The latter is $1$ in the case of a decoding failure, $0$ otherwise.
\item The analysis of the collected couples $\left\{ e^{(i)}, \Im^{(i)}\right\}$ provides the opponent with some information about the distance spectrum of the secret key.
\item The opponent exploits this information to reconstruct the secret key (or an equivalent representation of it).
\end{enumerate}
We point out that these attacks can affect code-based systems achieving security against both \ac{CPA} and \ac{CCA2}.
However, in this paper we only focus on systems with \ac{CCA2} security, which represent the most interesting case. Therefore, we assume that each decryption query uses an error vector randomly picked among all the $n$-tuples with weight $t$ (see Section \ref{sec:descryption}).

\subsection{Matrix reconstruction from the distance spectrum\label{sec:DSR}}

In \cite[section 3.4]{Fabsic2017} the problem of recovering the support of a vector from its distance spectrum has been defined as Distance Spectrum Reconstruction (DSR) problem, and can be formulated as follows:

\vspace{1em}
\textbf{Distance Spectrum Reconstruction (DSR)}
\\\textit{Given $\Delta(v)$, with $v$ being a $p$-bit vector with weight $w$, find a set of integers $\Theta^* = \left\{ v^*_0, v^*_1,\cdots,v^*_{w-1} \left|\hspace{2mm} v^*_i<p\right.\right\}$ such that $\Theta^*$ is the support of a $p$-bit vector $v^*$ and $\Delta(v^*)=\Delta(v)$}.
\vspace{1em}

This problem is characterized by the following properties:
\begin{itemize}
\item each vector obtained as the cyclic shift of $v$ is a valid solution to the problem; the search for a solution can then be made easier by setting $v^*_0=0$ and $v^*_1 = \min{\left\{ \left. \pm d \mod{p}\right|d\in \Delta(v) \right\}}$;
\item the elements of $\Theta^*$ must satisfy the following property:
\begin{equation}
\forall v^*_i > 0\hspace{2mm} 
\exists d\in\Delta(v) \text{\hspace{2mm}s.t.\hspace{2mm}}v^*_i = d \wedge v^*_i = p - d,
\end{equation}
since it must be $\delta(0,v^*_i) = \min{\left\{ v_i^*,p-v_i^*\right\}} \in \Delta(v)$.
\item for every solution $\Theta^*$, there always exists another solution $\Theta'$ such that:
\begin{equation}
\label{eq:compl_candidates}
\forall v^*_i\in\Theta^* \hspace{1mm}\exists! v'_i \in\Theta' \hspace{2mm}\text{s.t.}\hspace{2mm}v'_i=(p-v^*_i)\mod{p};
\end{equation}
\item the DSR problem can be represented through a graph $\mathcal{G}$, containing nodes with values $0,1,\cdots,p-1$: there is an edge between any two nodes $i$ and $j$ if and only if $\delta(i,j)\in\Delta(v)$.
In the graph $\mathcal{G}$, a solution $\Theta^*$ (and $\Theta'$) is represented by a size-$w$ clique.
\end{itemize}

Reaction attacks against \ac{QC-MDPC} code-based systems are based on the DSR problem. 
Instead, in the case of \ac{QC-LDPC} code-based systems, an attacker aiming at recovering the secret \ac{QC-LDPC} code has to solve the following problem:

\vspace{1em}
\textbf{Distance Spectrum Distinguishing and Reconstruction (DSDR)}\\\textit{Given $\bigcup_{i=0}^{z-1}\Delta\left(v^{(i)}\right)$, where each $v^{(i)}$ is a $p$-bit vector with weight $w^{(i)}$, find $z$ sets $\Theta^{*(i)} = \left\{ v^{*(i)}_0, v^{*(i)}_1,\cdots,v^{*(i)}_{w^{(i)}-1} \left|\hspace{2mm} v^{*(i)}_j<p\right.\right\}$ such that each $\Theta^{*(i)}$ is the support of a $p$-bit vector $v^{*(i)}$ and $\bigcup_{i=0}^{z-1}\Delta\left(v^{*(i)}\right)=\bigcup_{i=0}^{z-1}\Delta\left(v^{(i)}\right)$}.
\vspace{1em}

Also in this case, the problem can be represented with a graph, where solutions of the DSDR problem are defined by cliques of proper size and are coupled as described by \eqref{eq:compl_candidates}.
On average, solving these problems is easy: the associated graphs are sparse (the number of edges is relatively small), so the probability of having spurious cliques (i.e., cliques that are not associated to the actual distance spectrum), is in general extremely low.
In addition, the complexity of finding the solutions is significantly smaller than that of the previous steps of the attack, so it can be neglected \cite{Fabsic2017,Guo2016}.
From now on, we conservatively assume that these problems always have the smallest number of solutions, that is equal to $2$ for the DSR case and to $2z$ for the DSDR case.

\subsection{GJS attack}

The first reaction attack exploiting decoding failures has been proposed in \cite{Guo2016}, and is tailored to \ac{QC-MDPC} code-based systems.
Therefore, we describe it considering $Q=I_{n_0p}$, $\tilde{H} = H$, and we refer to it as the GJS attack.
In this attack, the distance spectrum recovery is performed through Algorithm \ref{alg:GJS}.
The vectors $a$ and $b$ estimated through Algorithm \ref{alg:GJS} are then used by the opponent to guess the multiplicity of each distance in the spectrum of $H_{n_0 - 1}$.
Indeed, the ratios $p_d = \frac{a_d}{b_d}$ follow different and distinguishable distributions, with mean values depending on the multiplicity of $d$.
This way, the analysis of the values $p_d$ allows the opponent to recover $\Delta\left(H_{n_0 - 1}\right)$.

\begin{algorithm}[ht!]
\caption{GJS distance spectrum recover}
\label{alg:GJS}
\hspace*{\algorithmicindent} 
\begin{algorithmic}
\State{$a\leftarrow$ zero initialized vector of length $\left\lfloor \frac{p}{2} \right\rfloor$}
\State{$b\leftarrow$ zero initialized vector of length $\left\lfloor \frac{p}{2} \right\rfloor$}
\For{$\left\{ i=0,1,\cdots,T-1\right\}$}
\State{$x^{(i)}\leftarrow$ ciphertext encrypted with the error vector $e^{(i)}$}
\State{Divide $e^{(i)}$ as $\left[e^{(i)}_0,\cdots,e^{(i)}_{n_0-1}\right]$, where each $e^{(i)}_j$ has length $p$}
\State{$\Delta (e^{(i)}_{n_0 - 1})\leftarrow$ distance spectrum of $e^{(i)}_{n_0-1}$}
\For{$\left\{ d \in \Delta\left(e^{(i)}_{n_0 - 1}\right)\right\}$}
\State{$b_d \leftarrow b_d+1$}
\State{$a_d \leftarrow a_d+\Im^{(i)}$}
\EndFor
\EndFor
\end{algorithmic}
\end{algorithm}

Solving the DSR problem associated to $\Delta\left(H_{n_0-1} \right)$ allows the opponent to obtain a matrix $H^*_{n_0-1} = \Pi H_{n_0-1}$, with $\Pi$ being an unknown circulant permutation matrix.
Decoding of intercepted ciphertexts can be done just with $H^*_{n_0-1}$. 
Indeed, according to \eqref{eq:public_key}, the public key can be written as $G'=[I | P]$, with:
\begin{equation}
\label{eq:matrixP}
P=\begin{bmatrix} P_0 \\
P_1 \\
\vdots \\
P_{n_0-2} \\ 
\end{bmatrix}=\begin{bmatrix} \left(H^{-1}_{n_0-1} H_{0}\right)^T \\
\left(H^{-1}_{n_0-1} H_{1}\right)^T \\
\vdots \\
\left(H^{-1}_{n_0-1} H_{n_0-2}\right)^T \\ 
\end{bmatrix}.
\end{equation}
The opponent can then compute the products:
\begin{equation}
H^*_{n_0-1}P_i^T = \Pi H_i = H_i^*,
\end{equation}
in order to obtain a matrix $H^* =\left[H^*_0, H^*_1,\cdots,H^*_{n_0-1}\right] = \Pi H$. 
This matrix can be used to efficiently decode the intercepted ciphertexts, since:
\begin{equation}
xH^{*T} = e H^T \Pi^T = s^T\Pi^T = s^{*T}.
\end{equation}
Applying a decoding algorithm on $s^{*T}$, with the parity-check matrix $H^*$, will return $e$ as output.
The corresponding plaintext can then be easily recovered by considering the first $k$ positions of $x+e$.

As mentioned in Section \ref{sec:DSR}, the complexity of solving the DSR problem can be neglected, which means that the complexity of the GJS attack can be approximated with the one of
Algorithm \ref{alg:GJS}.
First of all, we denote as $C_{\texttt{dist}}$ the number of operations that the opponent must perform, for each decryption query, in order to compute the distance spectrum of $e^{(i)}$ and update the estimates $a$ and $b$. 
The $p$-bit block $e^{(i)}_{n_0-1}$ can have weight between $0$ and $t$; let us suppose that its weight is $t_p$, which occurs with probability
\begin{equation}
p_{t_p} = \frac{\binom{p}{t_p}\binom{n-p}{t-t_p}}{\binom{n}{t}}.
\end{equation}
We can assume that in $e_{n_0-1}$ there are no distances with multiplicity $\geq 2$ (this is reasonable when $e$ is sparse). 
The average number of distances in $e_{n_0-1}$ can thus be estimated as $\sum_{t_p=0}^{t}{p_{t_p}\binom{t_p}{2}}$, which also gives the number of operations needed to obtain the spectrum of $e_{n_0-1}$.
Each of these distances is associated to two additional operations: the update of $b$, which is performed for each decryption query, and the update of $a$, which is performed only in the case of a decryption failure.
Thus, if we denote as $\epsilon$ the \ac{DFR} of the system and as $C_{\texttt{enc}}$ and $C_{\texttt{dec}}$ the complexities of one encryption and one decryption, respectively, the average complexity of each decryption query can be estimated as:
\begin{equation}
C_q = C_{\texttt{enc}}+C_{\texttt{dec}}+(2+\epsilon)\sum_{t_p=0}^{t}{p_{t_p}\binom{t_p}{2}}.
\end{equation}
Thus, the complexity of the attack, in terms of work factor, can be estimated as:
\begin{equation}
\label{eq:WF_GJS}
WF_{\texttt{GJS}}\approx T \cdot C_q = T \cdot \left[ C_{\texttt{enc}}+C_{\texttt{dec}}+(2+\epsilon)\sum_{t_p=0}^{t}{p_{t_p}\binom{t_p}{2}}\right].
\end{equation}

\subsection{FHS$^+$ attack}

More recently, a reaction attack specifically tailored to \ac{QC-LDPC} code-based systems has been proposed in \cite{Fabsic2017}, and takes into account the effect of the matrix $Q$.
We refer to this attack as the FHS$^+$ attack.
The collection phase in the FHS$^+$ attack is performed through Algorithm \ref{alg:FHSZGJ}.
We point out that we consider a slightly different (and improved) version of the attack in \cite{Fabsic2017}.
\begin{algorithm}[ht!]
\caption{FHS$^+$ distance spectrum recover}
\label{alg:FHSZGJ}
\hspace*{\algorithmicindent} 
\begin{algorithmic}
\State{$a\leftarrow$ zero initialized vector of length $\left\lfloor \frac{p}{2} \right\rfloor$}
\State{$b\leftarrow$ zero initialized vector of length $\left\lfloor \frac{p}{2} \right\rfloor$}
\State{$u\leftarrow$ zero initialized vector of length $\left\lfloor \frac{p}{2} \right\rfloor$}
\State{$v\leftarrow$ zero initialized vector of length $\left\lfloor \frac{p}{2} \right\rfloor$}
\For{$\left\{ i=0,1,\cdots,T-1\right\}$}
\State{$x^{(i)}\leftarrow$ ciphertext encrypted with the error vector $e^{(i)}$}
\For{$\left\{j=0,1,\cdots,n_0-1\right\}$}
\State{Divide $e^{(i)}$ as $\left[e^{(i)}_0,\cdots,e^{(i)}_{n_0-1}\right]$, where each $e^{(i)}_j$ has length $p$}
\State{$\Delta\left( e^{(i)}_{j} \right)\leftarrow$ distance spectrum of $e^{(i)}_{j}$}
\EndFor
\State{
$\Delta\left(e^{(i)}\right)=\bigcup_{j=0}^{n_0-1}\Delta\left( e^{(i)}_j \right)$}
\For{$\left\{ d \in \Delta\left(e^{(i)}\right)\right\}$}
\State{$b_d \leftarrow b_d+1$}
\State{$a_d \leftarrow a_d+\Im^{(i)}$}
\EndFor
\For{$\left\{ d \in \Delta\left(e^{(i)}_{n_0-1}\right)\right\}$}
\State{$v_d \leftarrow v_d+1$}
\State{$u_d \leftarrow u_d+\Im^{(i)}$}
\EndFor
\EndFor
\end{algorithmic}
\end{algorithm}

As in the GJS attack, the estimates $\frac{a_d}{b_d}$ are then used by the opponent to guess the distances appearing in the blocks of $H$.
In particular, every block $e^{(i)}_{j}$ gets multiplied by all the blocks $H_{j}$, so the analysis based on $\frac{a_d}{b_d}$ reveals $\Delta(H)=\bigcup_{j=0}^{n_0-1}{\Delta(H_{j})}$.
In the same way, the estimates $\frac{u_d}{v_d}$ are used to guess the distances appearing in the blocks belonging to the last block row of $Q^T$.
Indeed, the block $e^{(i)}_{n_0-1}$ gets multiplied by all the blocks $Q^T_{j,n_0-1}$.
Since a circulant matrix and its transpose share the same distance spectrum, the opponent is indeed guessing distances in the first block column of $Q$.
In other words, the analysis based on $\frac{u_d}{v_d}$ reveals  $\Delta(Q)=\bigcup_{j=0}^{n_0-1}{\Delta(Q_{j,n_0-1})}$.

The opponent must then solve two instances of the DSDR problem in order to obtain candidates for $H_j$ and $Q_{j,n_0-1}$, for $j=0,1,\cdots,n_0-1$. 
As described in Section \ref{sec:DSR}, we can conservatively suppose that the solution of the DSDR problem for $\Delta(H)$ is represented by two sets $\Gamma_h^{*}=\left\{ \Theta_h^{*(0)},\cdots,\Theta_h^{*(n_0-1)} \right\}$ and $\Gamma_h'=\left\{ \Theta_h'^{(0)},\cdots,\Theta_h'^{(n_0-1)} \right\}$, with each couple $\left\{ \Theta_h^{*(j)}, \Theta_h'^{(j)} \right\}$ satisfying \eqref{eq:compl_candidates}.
Each solution $\Theta^{*(j)}$ (as well as the corresponding $\Theta'^{(j)}$) represents a candidate for one of the blocks in $H$, up to a cyclic shift. In addition, we must also consider that the opponent has no information about the correspondence between cliques in the graph and blocks in $H$: in other words, even if the opponent
correctly guesses all the circulant blocks of $H$, he does not know their order and hence must consider all their possible permutations.
Considering the well-known isomorphism between $p \times p$ binary circulant matrices and polynomials in $GF_2[x]/(x^p+1)$, the matrix $H$ can be expressed in polynomial form as:
\begin{equation}
\label{eq:H_poly}
H(x) = \left[x^{\alpha_0^{(h)}}h_{\pi^{(h)}(0)}(x), x^{\alpha_1^{(h)}}h_{\pi^{(h)}(1)}(x),\cdots,x^{\alpha_{n_0-1}^{(h)}}h_{\pi^{(h)}(n_0-1)}(x) \right],
\end{equation}
with $\alpha_j^{(h)}\in[0 , p-1]$, $\pi^{(h)}$ being a permutation of $\left\{0,1,\cdots,n_0-1 \right\}$ (so that $\pi^{(h)}(j)$ denotes the position of the element $j$ in $\pi^{(h)}$),
and each $h_j(x)$ is the polynomial associated to the support $\Theta_h^{*(j)}$ or $\Theta_h'^{(j)}$.
In the same way, solving the DSDR problem for $\Delta(Q)$ gives the same number of candidates for the last column of $Q$, which are denoted as $q_{j,n_0-1}(x)$ in polynomial notation.
This means that for the last column of $Q(x)$ we have an expression similar to \eqref{eq:H_poly}, with $n_0$ coefficients $\alpha_j^{(q)}\in [0 , p-1]$ and a permutation $\pi^{(q)}$.
The opponent must then combine these candidates, in order to obtain candidates for the last block of the matrix $\tilde{H}=HQ$, which is denoted as $\tilde{h}_{n_0-1}(x)$ in polynomial form.
Indeed, once $\tilde{h}_{n_0-1}(x)$ is known, the opponent can proceed as in the GJS attack for recovering $\tilde{H}$.
Taking into account that $\tilde{H}=HQ$, the polynomial $\tilde{h}_{n_0-1}(x)$ can be expressed as:
\begin{equation}
\label{eq:h0_eq0}
\tilde{h}_{n_0-1}(x)=\sum_{j=0}^{n_0-1}{x^{\alpha^{(h)}_j} h_{\pi^{(h)}(j)}(x)x^{\alpha^{(q)}_j}q_{\pi^{(q)}(j),n_0-1}(x)}.
\end{equation}
Because of the commutative property of the addition, the opponent can look only for permutations of the polynomials $q_{j,n_0-1}(x)$.
Then, \eqref{eq:h0_eq0} can be replaced by:
\begin{equation}
\label{eq:h0_eq}
\tilde{h}_{n_0-1}(x)=\sum_{j=0}^{n_0-1}{x^{\alpha^{(h)}_j} h_j(x)x^{\alpha^{(q)}_j}q_{\pi^{(q)}(j),n_0-1}(x)},
\end{equation}
which can be rearranged as:
\begin{equation}
\tilde{h}_{n_0-1}(x)=\sum_{j=0}^{n_0-1}{x^{\alpha_j} h_j(x)q_{\pi^{(q)}(j),n_0-1}(x)},
\end{equation}
with $\alpha_j = \alpha^{(h)}_j+\alpha^{(q)}_j\mod{p}$.
Since whichever row-permuted version of $\tilde{H}$ can be used to decode intercepted ciphertexts, we can write:
\begin{align}
\label{eq:cand_hprime0}
\tilde{h}'_{n_0-1}(x)\nonumber&=x^{-\alpha_0}\tilde{h}_{n_0-1}(x)=\\\nonumber
&=x^{-\alpha_0}\sum_{j=0}^{n_0-1}{x^{\alpha_j} h_j(x)q_{\pi^{(q)}(j),n_0-1}(x)}=\\
&=\sum_{j=0}^{n_0-1}{x^{\beta_j} h_j(x)q_{\pi^{(q)}(j),n_0-1}(x)},
\end{align}
with $\beta_0 = 0$ and $\beta_j\in\left\{0, 1 , \cdots, p-1  \right\}$.

We must now consider the fact that, in the case of blocks $Q_{j,n_0-1}$ having weight $\leq 2$ (we suppose that the weights of the blocks $H_j$ are all $>2$), the number of candidates for $\tilde{h}'_{n_0-1}(x)$ is reduced.
Indeed, let us suppose that there are $n^{(2)}$ and $n^{(1)}$ blocks $Q_{j,n_0-1}(x)$ with weights $2$ and $1$, respectively. Let us also suppose that there is no null block in $Q$.
These assumptions are often verified for the parameter choices we consider.
For blocks with weight $1$ there is no distance to guess, which means that the associated polynomial is just $x^0=1$.
In the case of a block with weight $2$, the two possible candidates are in the form $x^0+x^d$ and $x^0+x^{p-d}$. 
However, since $x^0+x^{d} = x^d \left(x^0+x^{p-d} \right)$, the opponent can consider only one of the two solutions defined by \eqref{eq:compl_candidates}.

Hence, the number of possible choices for the polynomials $h_j(x)$ and $q_{j,0}(x)$ in \eqref{eq:cand_hprime0} is equal to $2^{2n_0-n^{(2)}-n^{(1)}}$.
In addition, the presence of blocks with weight $1$ reduces the number of independent configurations of $\pi^{(q)}$: if we look at \eqref{eq:h0_eq}, it is clear that any two permutations $\pi^{(q)}_1$ and $\pi^{(q)}_2$ 
that differ only in the positions of the polynomials with weight $1$ lead to two identical sets of candidates.
Based on these considerations, we can compute the number of different candidates in \eqref{eq:cand_hprime0} as:
\begin{equation}
N_c = \frac{n_0!}{n^{(1)}!} 2^{2n_0-n^{(1)}-n^{(2)}}p^{n_0-1}.
\end{equation}

The complexity for computing each of these candidates is low: indeed, the computations in \eqref{eq:cand_hprime0} involve sparse polynomials, and so they require a small number of operations.
For this reason, we neglect the complexity of this step in the computation of the attack work factor.
After computing each candidate, the opponent has to compute the remaining polynomials forming $\tilde{H}(x)$ through multiplications by the polynomials appearing in the non-systematic part of the public key (see \eqref{eq:public_key}).
In fact, it is enough to multiply any candidate for $\tilde{h}_{n_0-1}(x)$ by the polynomials included in the non-systematic part of $G'$ (see \eqref{eq:public_key}).
When the right candidate for $\tilde{h}_{n_0-1}(x)$ is tested, the polynomials resulting from such a multiplication will be sparse, with weight $\leq md_v$.
The check on the weight can be initiated right after performing the first multiplication: if the weight of the first polynomial obtained is $>md_v$, then the candidate is discarded, otherwise the other polynomials are computed and tested.
Thus, we can conservatively assume that for each candidate the opponent performs only one multiplication.
Considering fast polynomial multiplication algorithms, complexity can be estimated in $C_c = p\log_2(p)$.
Neglecting the final check on the weights of the vectors obtained, the complexity of computing and checking each one of the candidates of $\tilde{h}_{n_0-1}(x)$ can be expressed as:
\begin{equation}
\label{eq:new_FHSZGJ}
WF_{\texttt{FHS$^+$}}\geq N_c C_c = 2^{2n_0-n^{(1)}-n^{(2)}} \frac{n_0!}{n^{(1)}!}p^{n_0}\log_2(p).
\end{equation}
The execution of Algorithm \ref{alg:FHSZGJ} has a complexity which can be estimated in a similar way as done for the GJS attack (see eq. \eqref{eq:WF_GJS}).
However, unless the \ac{DFR} of the system is significantly low (such that $T$ is in the order of the work factor expressed by \eqref{eq:new_FHSZGJ}), collecting the required number of cyphertexts for the attack is negligible from the complexity standpoint \cite{Fabsic2017}, so \eqref{eq:new_FHSZGJ} provides a (tight) lower bound on the complexity of the attack.

\subsection{FHZ attack}
The FHZ attack has been proposed in \cite{Fabsic2018}, and is another attack procedure specifically tailored to \ac{QC-LDPC} code-based systems.
The attack starts from the assumption that the number of decryption queries to the oracle is properly bounded, such that the opponent cannot recover the spectrum of $H$ (this is the design criterion followed by the authors of LEDApkc \cite{LEDApkc}).
However, it may happen that such a bounded amount of ciphertexts is enough for recovering the spectrum of $Q$: in such a case, the opponent might succeed in reconstructing a shifted version of $H$, with the help of \ac{ISD}. 
The distance spectrum recovery procedure for this attack is described in Algorithm \ref{alg:FHZ}. 
\begin{algorithm}[ht!]
\caption{FHZ distance spectrum recovery}
\label{alg:FHZ}
\hspace*{\algorithmicindent} 
\begin{algorithmic}
\For{$\left\{j=0,1,\cdots,n_0-1\right\}$}
\State{$a^{(j)}\leftarrow$ zero initialized vector of length $\left\lfloor \frac{p}{2} \right\rfloor$}
\State{$b^{(j)}\leftarrow$ zero initialized vector of length $\left\lfloor \frac{p}{2} \right\rfloor$}
\EndFor
\For{$\left\{ i=0,1,\cdots,M\right\}$}
\State{$x^{(i)}\leftarrow$ ciphertext encrypted with the error vector $e^{(i)}$}
\State{Divide $e^{(i)}$ as $\left[e^{(i)}_0,\cdots,e^{(i)}_{n_0-1}\right]$, where each $e^{(i)}_j$ has length $p$}
\For{$\left\{j=0,1,\cdots,n_0-1\right\}$}
\State{$\Delta\left( e^{(i)}_{j} \right)\leftarrow$ distance spectrum of $e^{(i)}_{j}$}
\For{$\left\{ d \in \Delta\left(e^{(i)}_j\right)\right\}$}
\State{$b^{(j)}_d \leftarrow b_d+1$}
\State{$a^{(j)}_d \leftarrow a_d+\Im^{(i)}$}
\EndFor
\EndFor
\EndFor
\end{algorithmic}
\end{algorithm}
\\The estimates  $\frac{a_d^{(i)}}{b_d^{(i)}}$ are then used to guess distances in $\bigcup_{j=0}^{n_0-1}{\Delta\left( Q_{j,i}\right)}$; solving the related DSDR problems gives the opponent proper candidates for the blocks of $Q$.
These candidates can then be used to build sets of candidates for $Q^T$, which will be in the form:
\begin{equation}
\label{eq:cand_Q}
Q^T(x)=\begin{bmatrix} x^{\alpha_0} \tilde{q}_0(x) & x^{\alpha_{n_0}} \tilde{q}_{n_0}(x) & \cdots & x^{\alpha_{n_0(n_0 - 1)}} \tilde{q}_{n_0(n_0 - 1)}(x)\\
x^{\alpha_1} \tilde{q}_1(x) & x^{\alpha_{n_0+1}} \tilde{q}_{n_0+1}(x) & \cdots & x^{\alpha_{n_0(n_0 - 1)+1}} \tilde{q}_{n_0(n_0 - 1)+1}(x)\\
\vdots & \vdots & \ddots & \vdots\\
x^{\alpha_{n_0-1}} \tilde{q}_{n_0-1}(x) & x^{\alpha_{2n_0-1}} \tilde{q}_{2n_0-1}(x) & \cdots & x^{\alpha_{n^2_0 - 1}} \tilde{q}_{n^2_0 - 1}(x),
\end{bmatrix}
\end{equation}
where each polynomial $\tilde{q}_j(x)$ is obtained through the solution of the DSDR problem (in order to ease the notation, the polynomial entries of $Q^T(x)$ in \eqref{eq:cand_Q} have been put in sequential order, such that we can use only one subscript to denote each of them).
Let us denote as $\bar{m}=\left[ m_0,m_1,\cdots,m_{n_0-1}\right]$ the sequence of weights defining the first row of $w(Q)$, as explained in Section \ref{sec:descryption}.
The solutions of the DSDR problem for the first row of $Q^T$ will then give two polynomials for each weight. 
The number of candidates for $Q^T(x)$ depends on the distribution of the weights in $\bar{m}$: let us consider, for the sake of simplicity, the case of $m_0=m_1$, while all the other weights in $\bar{m}$ are distinct.
In this situation, the graph associated to the DSDR problem will contain (at least) two couples of cliques with size $m_0=m_1$ (see Section \ref{sec:DSR}).
For the sake of simplicity, let us look at the first row of $Q^T(x)$: in such a case, the solution is represented by the sets $\Gamma^{(1)} = \left\{\Theta^{*(1)}, \Theta'^{(1)} \right\}$ and $\Gamma^{(2)} = \left\{\Theta^{*(2)}, \Theta'^{(2)} \right\}$, where each couple of cliques $\Theta^{*(i)}$ and $\Theta^{'(i)}$ is described by \eqref{eq:compl_candidates}.
In order to construct a candidate for $Q^T(x)$, as in \eqref{eq:cand_Q}, the opponent must guess whether $\Gamma^{(1)}$ is associated to $\tilde{q}_0(x)$ (and $\Gamma^{(2)}$ is associated to $\tilde{q}_{n_0}(x)$) or to $\tilde{q}_1(x)$; then, he must pick one clique from each $\Gamma^{(i)}$.
The number of candidates for the first row of $Q^T(x)$ is hence $2^{n_0}2! = 2^{n_0}2$. 
Since there are $n_0$ rows, the number of possible choices for the polynomials in \eqref{eq:cand_Q} is then equal to $(2!)^{n_0}2^{n^2_0} = 2^{n_0}2^{n^2_0}$.
If we have $m_0=m_1=m_2$, then this number is equal to $(3!)^{n_0}2^{n^2_0}$. 
In order to generalize this reasoning, we can suppose that $\bar{m}$ contains $z$ distinct integers $\hat{m}_0,\hat{m}_1,\cdots,\hat{m}_{z-1}$, with multiplicities $j_0,j_1,\cdots,j_{z-1}$, that is:
\begin{equation}
j_i=\#_l\left\{m_l=\hat{m}_i \right\}.
\end{equation}
Thus, also taking into account the fact that for polynomials with weight $\leq 2$ we have only one candidate (instead of $2$), the number of different choices for the entries of $Q^T(x)$ in \eqref{eq:cand_Q} can be computed as:
\begin{equation}
\label{eq:NQ}
N_Q=2^{n^2_0-n_0n^{(2)}-n_0n^{(1)}} \left[ \prod_{\begin{smallmatrix}i=0\\ \hat{m}_i\geq 2\end{smallmatrix}}^{z}{j_i!}\right]^{n_0},
\end{equation}
with $n^{(2)}$ and $n^{(1)}$ being the number of entries of $\bar{m}$ that are equal to $2$ and $1$, respectively.
Considering \eqref{eq:cand_Q}, the $i$-th row of $Q^T$ can be expressed as $\bar{q}_i(x)S_i(x)$, with:
\begin{equation}
\bar{q}_i(x) = \left[\tilde{q}_i(x) , \tilde{q}_{n_0+i}(x) , \cdots , \tilde{q}_{n_0(n_0-1)+i}(x) \right],
\end{equation}
and $S_i(x)$ being a diagonal matrix:
\begin{equation}
\label{eq:Si}
S_i(x) = \begin{bmatrix}x^{\alpha_i} & & & &\\
& x^{\alpha_{n_0+1+i}}&&&\\
& & x^{\alpha_{2n_0+3+i}} &&\\
& & &\ddots&\\
& & & & x^{\alpha_{n_0(n_0-1)+i}}
\end{bmatrix}.
\end{equation}
\\Let $G'(x)$ denote the public key; then, the matrix $G''(x)=G'(x)Q^T(x)$ is a generator matrix of the secret code.
In particular, we have:
\begin{align}
G''(x)&\nonumber=G'(x)Q^T(x)=\\\nonumber
&=
\begin{bmatrix}\begin{matrix}1 & & &\\
                & 1 & &\\
                & & \ddots &\\
                & & & 1
                \end{matrix} &
                \left|
                \begin{matrix}g_0(x) \\
g_1(x)\\
\vdots\\
g_{n_0-2}(x)\end{matrix}\right.\end{bmatrix}\cdot\begin{bmatrix}\bar{q}_0(x)S_0(x)\\
\bar{q}_1(x)S_1(x)\\
\vdots
\\\bar{q}_{n_0-1}(x)S_{n_0-1}(x)
\end{bmatrix}=
\\
&=\begin{bmatrix}\bar{q}_0(x)S_0(x)+g_0(x)\bar{q}_{n_0-1}(x)S_{n_0-1}(x)\\
\bar{q}_1(x)S_1(x)+g_1(x)\bar{q}_{n_0-1}(x)S_{n_0-1}(x)\\
\vdots\\
\bar{q}_{n_0-2}(x)S_{n_0-2}(x)+g_{n_0-2}(x)\bar{q}_{n_0-1}(x)S_{n_0-1}(x)
\end{bmatrix},
\end{align}
where $g_i(x)$ denotes the polynomial representation of the circulant obtained as $\left( \tilde{H}_{n_0-1}\tilde{H}_i\right)^T$.
The multiplication of every row of $G''(x)$ by whichever polynomial returns a matrix which generates the same code as $G''(x)$.
In particular, we can multiply the first row of $G''(x)$ by $x^{-\alpha_0}$, the second row by $x^{-\alpha_1}$, and so on.
The resulting matrix can then be expressed as:
\begin{align}
G(x)&\nonumber = \begin{bmatrix}x^{-\alpha_0} & & &\\
& x^{-\alpha_1}&&\\
& & \ddots&\\
& &  & x^{-\alpha_{n_0-2}}
\end{bmatrix} \cdot G''(x)=\\
&=\begin{bmatrix}x^{-\alpha_0}\left[\bar{q}_0(x)S_0(x)+g_0(x)\bar{q}_{n_0-1}(x)S_{n_0-1}(x)\right]\\
x^{-\alpha_1}\left[\bar{q}_1(x)S_1(x)+g_1(x)\bar{q}_{n_0-1}(x)S_{n_0-1}(x)\right]\\
\vdots\\
x^{-\alpha_{n_0-2}}\left[\bar{q}_{n_0-2}(x)S_{n_0-2}(x)+g_{n_0-2}(x)\bar{q}_{n_0-1}(x)S_{n_0-1}(x)\right]
\end{bmatrix}.
\end{align}
Taking into account \eqref{eq:Si}, we can define:
\begin{align}
\label{eq:Si_star}
S^*_i(x) & \nonumber = x^{-\alpha_i}S_i(x)=\\
&=\begin{bmatrix}1 & & & &\\
& x^{\alpha_{n_0+1+i}-\alpha_i}&&&\\
& & x^{\alpha_{2n_0+3+i}-\alpha_i} &&\\
& & &\ddots&\\
& & & & x^{\alpha_{n_0(n_0-1)+i}-\alpha_i}
\end{bmatrix},
\end{align}
which holds for $i\leq n_0-2$.
We can now express $G(x)$ as:
\begin{equation}
G(x)=\begin{bmatrix}\bar{q}_0(x)S^*_0(x)+g_0(x)\bar{q}_{n_0-1}(x)x^{-\alpha_0}S_{n_0-1}(x)\\
\bar{q}_1(x)S^*_1(x)+g_1(x)\bar{q}_{n_0-1}(x)x^{-\alpha_{1}}S_{n_0-1}(x)\\
\vdots\\
\bar{q}_{n_0-2}(x)S^*_{n_0-2}(x)+g_{n_0-2}(x)\bar{q}_{n_0-1}(x)x^{-\alpha_{n_0-1}}S_{n_0-1}(x)
\end{bmatrix}.
\end{equation}
As anticipated, $G(x)$ is a generator matrix for the secret code, which means that it admits
$H(x)$ as a corresponding sparse parity-check matrix.
Then, any row of the binary matrix corresponding to $\left[h_0(x),h_1(x),\cdots,h_{n_0-1}(x)\right]$ is a low-weight codeword in the dual of the code generated by $G(x)$.
Thus, an opponent can apply an \ac{ISD} algorithm to search for vectors with weight $n_0d_v$, denoted as $\bar{v}(x)$ in polynomial notation, such that $G(x)\bar{v}^T(x)=0$.
Finding one of these vectors results in determining
a row of the secret parity-check matrix.

For every non-singular matrix $A(x)$, we can define $G^*(x)=G(x)A(x)$ and $\bar{w}(x)=\bar{v}(x)A^{-T}(x)$, such that:
\begin{align}
G^*(x)\bar{w}^T(x) \nonumber & = G(x)A(x)\bar{w}^T(x)=\\\nonumber
&=G(x)A(x)A^{-1}(x)\bar{v}^T(x)=\\
&=G(x)\bar{v}^T(x)=0.
\end{align}
The opponent can apply \ac{ISD} on $G^*(x)$, searching for solutions $\bar{w}(x)$, and then obtain the corresponding 
vectors $\bar{v}(x)$ as $\bar{v}(x)=\bar{w}(x)A^T(x)$.
In particular, he can choose $A(x) = S^{*-1}_0(x)$, that is:
\begin{equation}
\label{eq:S0_inv}
A(x)=\begin{bmatrix}1 & & & &\\
& x^{\alpha_0-\alpha_{n_0+1}}&&&\\
& & x^{\alpha_0-\alpha_{2n_0+3}} &&\\
& & &\ddots&\\
& & & & x^{\alpha_0-\alpha_{n_0(n_0-1)}}
\end{bmatrix}.
\end{equation}

Let us denote as $\bar{g}^*_i(x)$ the $i$-th row of $G^*(x)$, and as $\bar{g}_i(x)$ the $i$-th row of $G(x)$; we have:
\begin{equation}
\bar{g}^*_i(x) = \bar{g}_i(x) A(x),
\end{equation}
which can be expressed as:
\begin{equation}
\bar{g}^*_i(x)=\left[\bar{q}_i(x)S^*_i(x)+g_i(x)\bar{q}_{n_0-1}(x)x^{-\alpha_i}S_{n_0-1}(x)\right]S^{*-1}_0(x). \end{equation}
For the first row of $G^*(x)$ (i.e., $i=0$), we have:
\begin{align}
\label{eq:g0}
\bar{g}^*_0(x)&\nonumber=\left[\bar{q}_0(x)S^*_0(x)+g_0(x)\bar{q}_{n_0-1}(x)x^{-\alpha_0}S_{n_0-1}(x)\right]S^{*-1}_0(x)=\\\nonumber
&=\bar{q}_0(x)+g_0(x)\bar{q}_{n_0-1}(x)x^{-\alpha_0}S_{n_0-1}(x)S^{*-1}_0(x)=\\
&=\bar{q}_0(x)+g_0(x)\bar{q}_{n_0-1}(x)D(x),
\end{align}
with $D(x) = x^{-\alpha_0}S_{n_0-1}(x)S^{*-1}_0(x)$ being a diagonal matrix with monomial entries only.
Once the polynomials $\bar{q}_0(x)$ and $\bar{q}_{n_0-1}$ have been picked, $\bar{g}^*_0(x)$ depends only on the values of the matrix $D(x)$.
This results in $p^{n_0}$ possible different candidates for $\bar{g}^*_0(x)$.

We can now look at the other rows of $G$; in general, the $i$-th row (with $i\geq 1$) is in the form:
\begin{align}
\label{eq:g_star}
\bar{g}^*_i(x)&\nonumber=\left[\bar{q}_i(x)S^*_i(x)+g_i(x)\bar{q}_{n_0-1}(x)x^{-\alpha_i}S_{n_0-1}(x)\right]S^{*-1}_0(x)=\\\nonumber
&=\bar{q}_i(x)S^*_i(x)S^{*-1}_0(x)+g_i(x)\bar{q}_{n_0-1}(x)x^{-\alpha_i}S_{n_0-1}(x)S^{*-1}_{0}(x)=\\
&=\bar{q}_i(x)S^*_i(x)S^{*-1}_0(x)+g_i(x)\bar{q}_{n_0-1}(x)x^{\alpha_0-\alpha_i}D(x).
\end{align}
From \eqref{eq:g_star} we see that the row $\bar{g}^*_i(x)$ is defined by $n_0$ independent parameters: indeed, 
$S^*_i(x)S^{*-1}_0(x)$ always has the first element equal to $1$, with all the other $n_0-1$ ones taking values in $[0 , p-1]$, 
while the only other additional degree of freedom comes from the choice of $(\alpha_0-\alpha_i)\in[0 , p-1]$. 

Based on the above considerations, we can finally obtain the total number of candidates for $G^*(x)$: starting from a choice of polynomials 
$\tilde{q}_0(x),\cdots,\tilde{q}_{n^2_0-1}(x)$, the opponent has $p^{n_0}$ independent possible choices for each row of $G^{*}(x)$.
Since the matrix has $n_0-1$ rows, the total number of candidates for $G^*(x)$ is then equal to:
\begin{equation}
\label{eq:NG}
N_G = \left( p^{n_0} \right)^{n_0-1}=p^{n^2_0-n_0}.
\end{equation}
For each candidate of $G^*(x)$, the opponent performs \ac{ISD}, searching for vectors $\bar{w}(x)$.
Since $A(x)$ is a permutation, the weight of $\bar{w}(x)$ is equal to that of $\bar{v}(x)$. 
Thus, the complexity of this last step is equal to that of \ac{ISD} running on a code with length $n=n_0p$, dimension $p$ (the opponent attacks the dual of the code generated by $G^*(x)$), searching for a codeword with weight $n_0d_v$, and can be denoted as $C_{\texttt{ISD}}\left( n_0p,p,n_0d_v\right)$.

As for the FHS$^+$ attack, unless the \ac{DFR} of the system is significantly low, we can neglect the complexity of Algorithm \ref{alg:FHZ}, and estimate the complexity of the FHZ attack as:
\begin{equation}
\label{eq:WF_FHZ}
WF_{\texttt{FHZ}} = N_Q\cdot N_G\cdot C_{\texttt{ISD}}\left( n_0p,p,n_0d_v\right),
\end{equation}
where $N_Q$ and $N_G$ are given by \eqref{eq:NQ} and \eqref{eq:NG}, respectively.

\section{Efficiency of reaction attacks \label{sec:further_cons}}
In the previous sections we have described reaction attacks against \ac{QC-MDPC} and \ac{QC-LDPC} code-based McEliece cryptosystems. 
In particular, we have computed the number of candidates an opponent has to consider for general choices of the system parameters, and devised tight complexity estimations.
Based on the analysis developed in the previous sections, in this section we study and compare the efficiency of all the aforementioned attack procedures.
First of all, we must consider the fact that the GJS attack can be applied to a \ac{QC-LDPC} code-based cryptosystem as well, on condition that Q-decoding is used for decryption.
In fact, as explained in section \ref{sec:Qdecoder}, the Q-decoder approximates a BF decoder working in $\tilde{H}$, therefore an attacker could focus on $\tilde{H}$ as the target of a reaction attack.

\begin{figure}[ht]
\centering
\includegraphics[width=95mm,keepaspectratio]{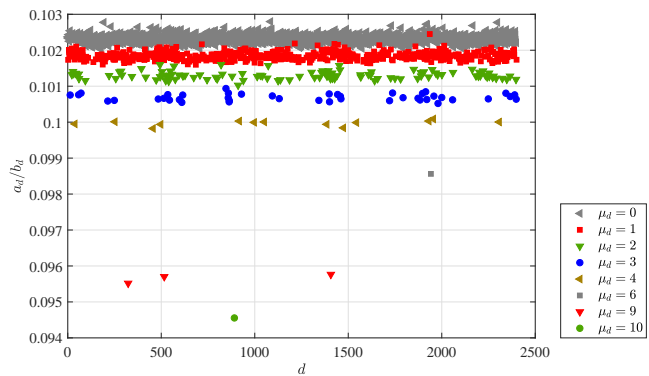}
\caption{Distribution of the opponent's estimates for a \ac{QC-LDPC} code-based system instance with $n_0=2$, $p=4801$, $d_v=9$, $[m_0,m_1]=[2,3]$, $t=95$, decoded through the Q-decoder. 
The estimates $a_d / b_d$ correspond to the output of Algorithm \ref{alg:GJS}.}
\label{fig:p4801_HQ}
\end{figure}

Based on this consideration, we can expect the GJS attack to be successful when Q-decoding is used: in such a case, the recovered distance spectrum is that of $\tilde{H}_{n_0-1}$ (see eq. \eqref{eq:Htilde}).
In order to verify this intuition, we have simulated the attack on a code with parameters $n_0=2$, $p=4801$, $d_v = 9$, $m=5$.
The corresponding estimates $\frac{a_d}{b_d}$, obtained through Algorithm \ref{alg:GJS}, are shown in Fig. \ref{fig:p4801_HQ}. 
As we can see from the figure, the distances tend to group into distinct bands, depending on the associated multiplicity in the 
spectrum.
\begin{figure}[ht]
\centering
\includegraphics[width=95mm,keepaspectratio]{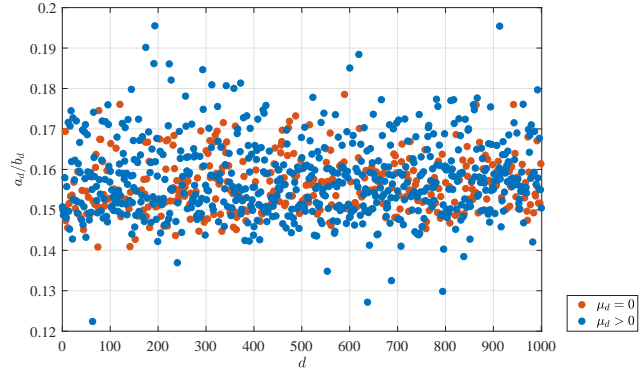}
\caption{Distribution of the opponent's estimates for a \ac{QC-LDPC} code-based system instance with $n_0=3$, $p=2003$, $d_v=7$, $[m_0,m_1,m_2]=[3,2,2]$, $t=12$, decoded through BF decoding working on the private \ac{QC-LDPC} code. 
The estimates $a_d / b_d$ correspond to the output of Algorithm \ref{alg:GJS}. 
}
\label{fig:p2003_HQ}
\end{figure}
In this case, the opponent can reconstruct the matrix $\tilde{H}_{n_0-1}$ by solving the related DSR problem, searching for cliques of 
size $\leq md_v$.

Instead, the same attack cannot be applied against a \ac{QC-LDPC} code-based system instance if BF decoding working on the private code is exploited.
In order to justify this fact, let us consider the generic expression of a block of $\tilde{H}$, say the first one:
\begin{equation}
\label{eq:h_tilde}
\tilde{h}_0(x)=
\sum_{j=0}^{n_0-1}{q_{j,0}(x)h_j(x)}=\sum_{j=0}^{n_0-1}a_j(x),
\end{equation}
with $a_j = q_{j,0}(x)h_j(x)$.
Each $a_j(x)$ can be seen as a sum of replicas of $h_j(x)$ (resp. $q_{j,0}(x)$) placed at positions depending on $q_{j,0}(x)$ (resp. $h_j(x)$).
Since all these polynomials are sparse, the expected number of cancellations occurring in such a sum is small.
This means that, with high probability, distances in $q_{j,0}(x)$ or $h_j(x)$ are present also in $a_j(x)$.
Since the BF decoder performance depends on distances in both $H$ and $Q$, the opponent can correctly identify these distances by analyzing Bob's reactions.
However, the spectrum of $\tilde{h}_0(x)$ also contains a new set of \textit{inter-block} distances, i.e., distances formed by one entry of $a_i(x)$ and one entry of $a_j(x)$, with $i\neq j$.
These distances cannot be revealed by the opponent, because they do not affect the decoding performance when a BF decoder working on the private code is used.
To confirm this statement, an example of the opponent estimates, obtained though Algorithm \ref{alg:GJS} for a \ac{QC-LDPC} code-based system instance exploiting BF decoding over the private \ac{QC-LDPC} code, is shown in Fig. \ref{fig:p2003_HQ}. From the figure we notice that, differently from the previous case, the two sets of distances are indistinguishable.

We can now sum up all the results regarding reaction attacks against the considered systems, and this is done in Table \ref{tab:reaction_attacks}, where the applicability of each attack against each of the considered systems is summarized, together with the relevant complexity.

\begin{table}[!t]
\caption{Applicability of reaction attacks, for different McEliece variants
\label{tab:reaction_attacks}}
\centering
\begin{tabular}{|c||c|c|c|c|}
\hline 
\textbf{Attack} & \textbf{Complexity} & \textbf{QC-MDPC} & $\begin{matrix}\textbf{QC-LDPC}\\
\textbf{(Q-decoder)}\end{matrix}$ & $\begin{matrix}\textbf{QC-LDPC}\\
\textbf{(BF decoder)}\end{matrix}$\\\hline\hline
\textbf{GJS} & Eq. \eqref{eq:WF_GJS} & \checkmark & \checkmark  & \xmark \\\hline 
\textbf{FHS$^+$} & Eq. \eqref{eq:new_FHSZGJ} & \xmark & \checkmark & \checkmark  \\\hline
\textbf{FHZ} &  Eq. \eqref{eq:WF_FHZ}& \xmark & \checkmark & \checkmark  \\\hline
\end{tabular}
\end{table}

The \ac{QC-MDPC} code-based system and the \ac{QC-LDPC} code-based system with Q-decoding are both exposed to the GJS attack.
For these systems, such an attack can be avoided only by achieving sufficiently low \ac{DFR} values, which is a solution that obviously prevents all reaction attacks.
Another solution consists in properly bounding the lifetime of a key-pair, which means that the same key-pair is used only for a limited amount of encryptions/decryptions, before being discarded.
Basically, this is equivalent to assume that the opponent can only exploit a bounded number of decryption queries.
The most conservative choice consists in using ephemeral keys, i.e., refreshing the key-pair after decrypting each ciphertext \cite{BIKE,LEDAkem}.
This choice allows avoiding reaction attacks of any type, but necessarily decreases the system efficiency.
Relaxing this condition would obviously be welcome, but estimating a safe amount of observed ciphertexts might be a hard task. 
A less drastic but still quite conservative choice might be bounding the lifetime of a key-pair as DFR$^{-1}$ (this means that, on average, the opponent has only one decryption query for each key-pair).
Actually, recent proposals achieve \ac{DFR} values in the order of $10^{-9}$ or less \cite{LEDAkem}, resulting into very long lifetimes for a key-pair.

When we consider classical BF decoding in the \ac{QC-LDPC} code-based system, the scenario is different.
In such a case, for a non-negligible DFR, we have to consider the complexities of both FHS$^+$ and FHZ attacks. 
Since for both attacks we have a precise estimation of the complexity, we can choose proper parameters to achieve attack work factors that are above the target security level.
\begin{table}[!t]
\caption{Sets of parameters of \ac{QC-LDPC} code-based system instances using BF decoding on the private code and achieving a security level of $2^{80}$ or more against FHS$^+$, FHZ and \ac{ISD} based attacks.
\label{tab:sys_parameters80}}
\centering
\begin{tabular}{|c|c|c|c|c|}
\hline  
$\mathbf{n_0}$ & $\mathbf{d_v}$ & $\mathbf{p}$ & $\mathbf{\bar{m}}$ & $\mathbf{t}$  \\\hline\hline
5 & 9 & 8539 & $[3,3,2,2,1]$ & 38\\\hline
5 & 9 & 7549 & $[3,2,2,1,1]$ & 37\\\hline
6 & 9 & 5557 & $[3,2,1,1,1,1]$ & 34\\\hline
6 & 11 & 5417 & $[2,1,1,1,1,1]$ & 34\\\hline
\end{tabular}
\end{table}
\begin{table}[!t]
\caption{Sets of parameters of \ac{QC-LDPC} code-based system instances using BF decoding on the private code and achieving a security level of $2^{128}$ or more against FHS$^+$, FHZ and \ac{ISD} based attacks.
\label{tab:sys_parameters128}}
\centering
\begin{tabular}{|c|c|c|c|c|}
\hline  
$\mathbf{n_0}$ & $\mathbf{d_v}$ & $\mathbf{p}$ & $\mathbf{\bar{m}}$ & $\mathbf{t}$  \\\hline\hline
$8$ & 9 & $13367$ & $[2,2,2,2,2,1,1,1]$ & 45\\\hline
8 & 11 & 14323 & $[2,2,2,2,2,1,1,1]$ & 44\\\hline
9 & 9 & 10657 & $[2,2,1,1,1,1,1,1,1]$ & 42\\\hline
9 & 11 & 11597 & $[2,2,1,1,1,1,1,1,1]$ & 42 \\\hline
\end{tabular}
\end{table}
In Tables \ref{tab:sys_parameters80} and \ref{tab:sys_parameters128} we provide some parameter choices able to guarantee that both reaction and \ac{ISD} attacks have a complexity of at least $2^{80}$ and $2^{128}$ operations, respectively.
We point out that, when $n_0$ increases, satisfying the conditions that ensure non-singularity of $Q$ according to Section \ref{key:gen} is no longer possible.
However, these conditions are sufficient but not necessary.
This means that, in some cases, the generation of $Q$ should be repeated, until a non-singular matrix is obtained.
We point out that the use of the BF decoder obviously leads to an increase in the code length (with respect to the Q-decoder), and this is the price to pay for withstanding reaction attacks.

\section{Conclusion}
In this paper we have analyzed recent reaction attacks against McEliece cryptosystems based on iteratively decoded codes.
We have generalized the attack procedures for all possible system variants and parameter choices, and provided estimates for their complexity.

For \ac{QC-MDPC} code-based systems, preventing reaction attacks requires achieving negligible \ac{DFR}, and the same occurs for \ac{QC-LDPC} code-based systems exploiting Q-decoding.
However, in the case of \ac{QC-LDPC} code-based systems, such attacks can be made infeasible by using the BF decoder and choosing proper parameters. 
This choice comes with the inevitable drawback of increasing the public key size, since the BF decoder is characterized by a worse performance than the Q-decoder.

In our analysis we have neglected the fact that, in all the attacks against \ac{QC-LDPC} code-based systems using BF decoding, the opponent must solve instances of the DSDR problem.
This problem can be made more difficult by appropriately choosing the distances in the spectrum of $Q$.
In other words, we can choose the blocks of $Q$ such that the union of the spectra $\bigcup_{j=0}^{n_0-1}{\left( Q_{j,n_0-1}\right) }$ forms a clique having size larger than the maximum value appearing in $\bar{m}$.
In this case, the number of solutions to the DSDR problem should be significantly increased.
This, however, is left for future works.

\section*{Acknowledgment}
The authors wish to thank Tom\'{a}\v{s} Fab\v{s}i\v{c} for fruitful discussion about the FHZ attack.

\bibliographystyle{splncs03}
\bibliography{Archive}

\end{document}